# Neutron Single-particle States in $^{101}$Sn by Polynomial Fits and Shell Model Calculations for Light Sn Isotopes

Abderrahmane Yakhelef[1,2] and Serkan Akkoyun[3]

[1]Physics department, faculty of sciences, Farhat Abbas Setif1 University, Algeria

[2]PRIMALab laboratory, El-Hadj Lakhedar Batna 1 University, Batna, Algeria

[3]Department of Physics, Faculty of Science, Sivas Cumhuriyet University Sivas, Turkey

**Abstract**

One of the main ingredients in nuclear structure studies using shell model are the single-particle energy (spe). In order to obtain these values accurately, experimental data is needed. The region around the doubly magic nuclide $^{100}$Sn is very interesting for nuclear studies in terms of structure, reaction and nuclear astrophysics. Experimental spectrum data for the $^{101}$Sn isotope is required for nuclear shell model studies to be carried out in this region. Since there is not enough experimental data in the literature, different approaches are used for the obtaining spe's for the region such as using the hole excitation spectrum in $^{131}$Sn or using the lightest and closest isotope $^{107}$Sn which figures the model space orbitals. In this work, we have performed second order polynomial fits of the tree single-particle states $s_{1/2}$, $d_{3/2}$ and $h_{11/2}$ in the light Sn isotopes up to $^{113}$Sn and $^{115}$Sn which are not determined yet experimentally. By an extrapolation toward light Sn isotopes, we can obtain the excitation energies of all the single-particle states in $^{101}$Sn. Subsequently, neutron spe's of the model space orbitals are defined. Shell model calculations for even and odd $^{102-107}$Sn isotopes are carried out using the new interactions and the results are compared with the experimental data and results obtained using the widely used interaction sn100pn.

1. Introduction

In the investigations of neutron deficient nuclei far from the beta stability line and close to the proton dripline, $^{100}$Sn isotope region is one of the unique choices. The heaviest double magic self-conjugate $^{100}$Sn isotope is very attractive for many reasons such as shell evolution, change of collective properties, band termination, and magnetic rotation. The use of high-power detector arrays and radioactive ion beams opens the opportunity to perform experimental studies in this region. By increasing the experimental excited energy information in this region, the results of the studies to be carried out with the theoretical models get closer to real experimental values. This allows more accurate approaches to be performed in nuclear structure studies and enables models to be examined with higher accuracy [1-3].

In the $^{100}$Sn region, there is not enough experimental excited state energy information in the literature to obtain single-particle energy (spe) values with great accuracy needed for the nuclear shell model calculations. Different approaches have been used in the calculations to obtain neutron spe values due to the lack of enough experimental data for the $^{101}$Sn isotope. Yakhelef and Bouldjedri [4] obtained the neutron single-particle energy values by using the excited energy states of the closest odd $^{107}$Sn isotope whose experimental data are available in the literature. Brown et al. [5] obtained neutron spe's by using the experimental energy spectrum of the $^{131}$Sn isotope. The interaction called *sn100pn* in that study was derived from a realistic interaction developed starting from the G-matrix derived from the CD Bonn nucleon-nucleon interaction. Hosaka et al. [6] is derived a set of interaction named as *sent* from a bare G-matrix based on the renormalized Paris potential for N=82 nuclei. The spe's of *sn100pn* and *snet* interactions are commonly used for the shell-model calculations performed on the $^{100}$Sn region. Trivedi et al. [7] modified these neutron spe's in the *sn100pn* and *snet* with the use of the value of 7/2$^+$ level of the $^{101}$Sn isotope which is the only experimental data available for this isotope. Leander et al. [8] theoretically obtained neutron spe's, based on Hartree-Fock with Skyrme III interaction, folded Yukawa potential and Wood-Saxon single-particle potentials, separately. Engeland et al. [9] modified the previous work mentioned by using experimental observations. Andreozzi et al. [10] resorted to the analysis of low energy spectra of isotopes with A<111, since there is not enough information in the literature regarding the spectrum of the $^{101}$Sn isotope. Sandulescu et al. [11] obtained neutron spe's by fitting known one quasi-particle excitation at $^{111}$Sn. Grawe2 et al. [12] and Schubart et al. [13] used $^{88}$Sr or $^{90}$Zr isotopes to theoretically obtain neutron spe's for this region.

Due to the lack of the experimental data on the $^{101}$Sn isotope in the literature, we have motivated to accurately obtain neutron spe's for $^{101}$Sn isotope. For the first time, we aimed to address the problem with a completely different approach from those previous studies in the literature. In the approach performed in the present study, we used 2$^{nd}$ order polynomial fits, for each set of experimental data, and make a smooth extrapolation toward the neutron dripline. Our goal is to obtain the excited energy spectrum of the $^{101}$Sn isotope, which allows us to deduce neutron spe's to be used in the nuclear shell-model calculations. Therefore, what we need to do is obtain the values of the 5/2$^+$, 7/2$^+$, 1/2$^+$, 3/2$^+$ and 11/2$^-$ excited energy levels of the $^{101}$Sn isotope. It has been observed that the results we obtained from the calculations performed with the spe values of polynomial fits are closer to the experimental data than the other calculations, especially for low-lying energy levels.

2. **Shell Model Calculations**

The nuclear shell model is one of the most suitable tools to describe the low-energy spectra of atomic nuclei [14-17]. Its main idea consists of considering the nucleus as a quantum system composed of A nucleons ( Z protons and N neutrons)  moving freely in a self-generated mean field. The Hamiltonian of the system can be written as in Eq. (1)

$$H = \sum_{i=1}^{A} T_i + \frac{1}{2}\sum_{i,j=1}^{A} V_{ij} \qquad (1)$$

Here $T_i$ is the kinetic energy of each nucleon, and $V_{ij}$ is the interaction potential between nucleons.  In order to overtake the complexity of the problem, an auxiliary one-body potential $U_i$ is introduced due to the fact that each nucleon moves under an average potential created by all the other nucleons. The Hamiltonian can be arranged as given in Eq. (2) and (3).

$$H = \sum_{i=1}^{A}[T_i + U_i] + \left[\frac{1}{2}\sum_{i,j=1}^{A} V_{ij} - \sum_{i,=1}^{A} U_i\right] \qquad (2)$$

$$H = H_0 + H_{res} \qquad (3)$$

Where, $H_0$ is the one-body Hamiltonian which describes the independent motion of the nucleons, and $H_{res}$ which represents the interaction Hamiltonian.

By using an average central potential to which a strong spin-orbit term is added, the single particle potential generates an energy spectrum organized in shells which explains the stable nucleon configurations corresponding to so determined magic numbers [14].

In the shell model frame the nucleus is composed of an inert core, made up of completely filled neutron and proton shells, plus n valence nucleons moving in a truncated model space and interacting through a model space effective interaction. The model space is spanned in general by a single major proton shell and/or a single major neutron shell above the inert core.

In any shell-model calculations, one has to start by defining a model space which is a set of active single-particle orbits outside the inert core. The basic inputs are the single particle matrix elements (spe) and the two-body matrix elements (tbme). For the former, spe are explicitly calculated using the mean-field models or defined empirically from the available experimental data of nuclei in the direct vicinity of the doubly magic nuclei. For the latter, the (tbme) are specified in terms of matrix elements of the residual interaction Hres, $< j1\ j2J|Hres|\ j3\ j4J >$, for all possible combinations of ji orbitals in the model space. J is the total two particles' angular momentum.

In the present study, we used the sn100pn interaction obtained by Brown et al [21] as the basis of this work. It is defined by a set of 862 two-body matrix elements, and the spe values have been determined using the whole excitation spectrum of $^{131}$Sn. The aim is to obtain more appropriate interaction for mass region around $^{100}$Sn by using single energy spectrum of $^{101}$Sn obtained by second order polynomial fits.

The final step in carrying out shell-model calculations is to diagonalize the model space-effective interaction with an appropriate algorithm such as Lanczos [20]. In this study, we used one of the well-known codes, the NuShellX@MSU code [18], to carry out the shell model spectroscopic calculations. It presents a development of NuShell code which contains a set of computer codes written by Brown and Rae [19]

### 3. Results and Discussions

*3.1 Obtaining spe values*

The experimental data of the single neutron states in light odd-Sn nuclei up to $^{115}$Sn are taken from [22] and shown in Fig1. We can consider $^{115}$Sn as a benchmark for the trend evolution of single particle states in odd Sn isotopes. Almost the five states start to change their direction of variation going forward to the neutron and proton drip-lines. States which are down in the region near $^{132}$Sn, $d_{3/2}$, $h_{11/2}$ and $s_{1/2}$, go up when going forward to the neutron drip-line near $^{100}$Sn. In contrast, the other states, $d_{5/2}$ and $g_{7/2}$, which are up near $^{132}$Sn go down to the ground-state and the first excited state in the other side of the nuclear chart.

The lightest isotope which figures all states is $^{109}$Sn. The state $s_{1/2}$ is still missing in the $^{107}$Sn nucleus spectrum. By going down to $^{101}$Sn, only the ground-state and the first excited state, believed to be $d_{5/2}$ and $g_{7/2}$ respectively, are defined experimentally. The energy splitting between these two levels in this nucleus has been reported in two different measurements as 172 keV [23,24].

In order to obtain the excited states in $^{101}$Sn which are not experimentally defined, we used a second order polynomial fit for each set of data corresponding to single-particle states $d_{3/2}$, $h_{11/2}$ and $s_{1/2}$ in odd Sn isotopes.

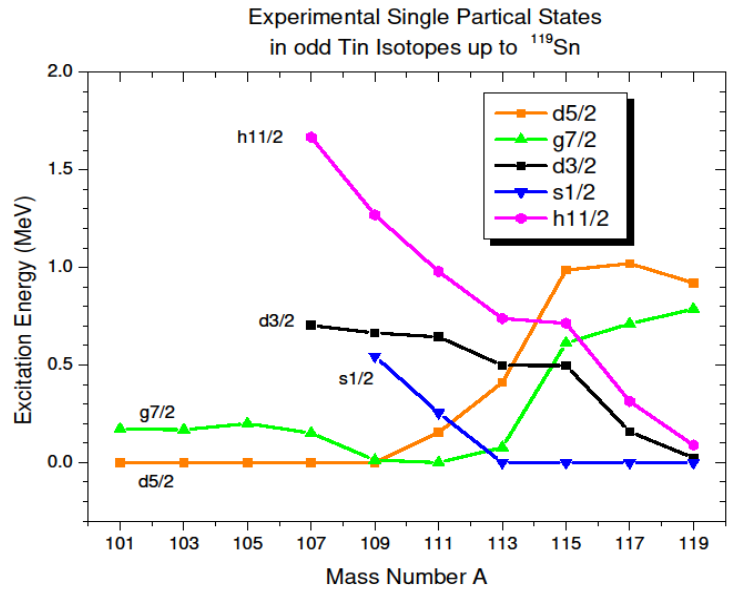

**Fig.1** Experimental single-particle states in odd Sn isotopes up to $^{119}$Sn

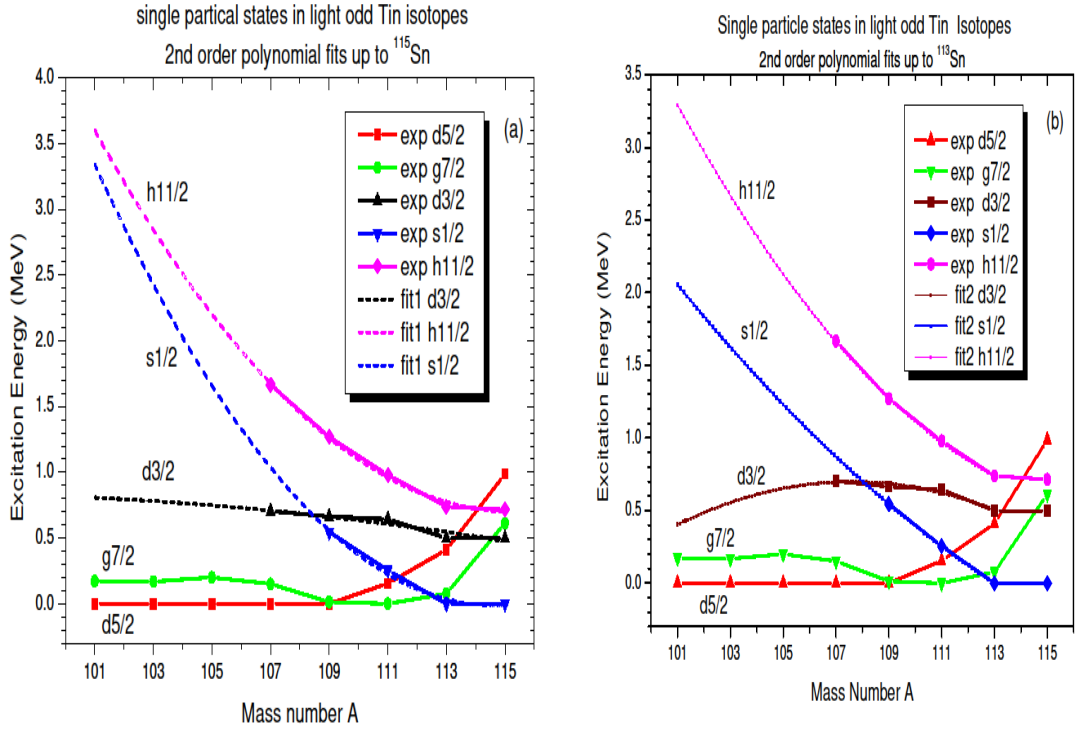

**Fig.2** Single-particle states in light Sn isotopes including (a) second order polynomial fit up to $^{115}$Sn (fit 1) and (b) second order polynomial fit up to $^{113}$Sn (fit 2)

We used two approaches based on the trend of the evolutions of states. In the first one, we took for each state of the experimental data in odd Sn isotopes up to $^{115}$Sn (fit 1). In the second approach we took only the data in odd Sn isotopes up to $^{113}$Sn (fit 2). The choice of the former is because, as mentioned above, the point where the trends of the evolution of states change between $^{100}$Sn and $^{132}$Sn. About the later, it looks like $^{113}$Sn is the point where the curves take a smooth evolution toward the proton drip-line. Results of the fits are represented in Fig.2 for these two different fits. By extrapolating curves down forward to lighter isotopes, we expect the excitation energies of these states in $^{101}$Sn with only one neutron outside the doubly magic inert core. Subsequently, we can calculate the corresponding spe's of the model space orbitals for each fit. Results are summarized in the Table 1.

**Table 1** Excitation and single-particle energies from the polynomial fits

|  | Single-particle energy (MeV) | | | | |
| --- | --- | --- | --- | --- | --- |
|  | $d_{5/2}$ | $g_{7/2}$ | $d_{3/2}$ | $s_{1/2}$ | $h_{11/2}$ |
| $E_{ex}$ (fit 1) | 0.000 | 0.172 | 0.811 | 3.347 | 3.605 |
| spe (fit 1) | -11.081 | -10.909 | -10.270 | -7.734 | -7.476 |
| $E_{ex}$ (fit 2) | 0.000 | 0.172 | 0.405 | 2.054 | 3.293 |
| spe (fit 2) | -11.081 | -10.909 | -10.676 | -9.027 | -7.788 |
| spe of sn100pn | -10.2893 | -10.6089 | -8.7167 | -8.6944 | -8.8152 |

*3.2 Shell model calculations*

Using the neutron spe values that we obtained in the present study from the two different second order polynomial fits (fit 1 and fit2), we performed nuclear shell model calculations to obtain energy spectrums for $^{102-107}$Sn isotopes. Besides, similar calculations using the widely used interaction sn100pn are carried out for the same nuclei. All results are compared to the existing experimental data and to each other.

Results of shell model calculations of $^{102}$Sn nucleus, of which few experimental data are available, are given in Fig. 3. As it is clearly seen, the three interactions reproduce correctly the spin and parity of the ground state. For the first excited $2^+$ state, it is clear that the two interactions fit1 and fit2 reproduce this state better than sn100pn. The closest one to the experimental data is fit2. For the two remaining states in the experimental spectrum $4^+$ and $6^+$, sn100pn reproduces these two states in invers order where $6^+$ was in good position compared to experiment but $4^+$ is higher to $6^+$. In contrast, we obtain these two states in their correct order using the neutron spe by fit 1 and fit 2. Fit 1 reproduces them lower by 170 KeV to the experiment while maintaining the energy difference between them. Fit 2 reproduces $6^+$ in its exact level, while $4^+$ is a bit lower by 215 keV compared to the experiment. In general, the two obtained interactions are better compared to the sn100pn. Fit 2 do it very good as it reproduces exactly the experimental states to an approximation of a few tens of keV, except $4^+$ state.

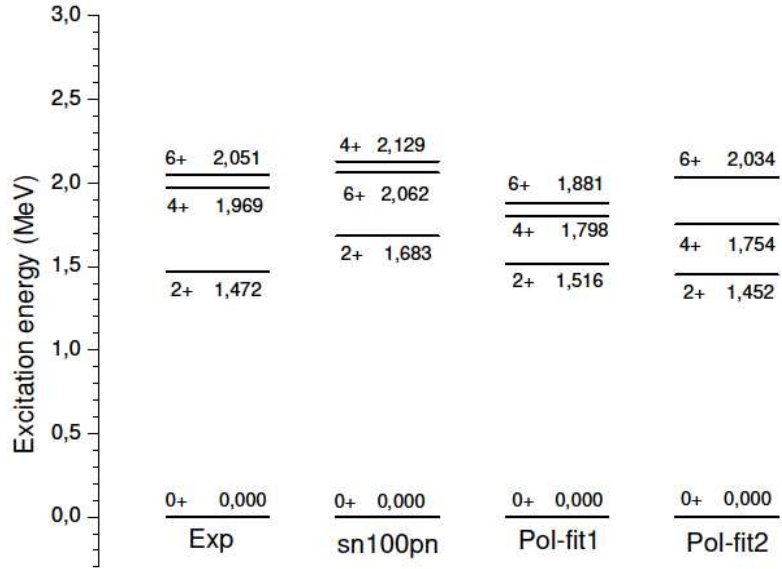

**Fig. 3.** Experimental and calculated low laying states of 102Sn isotope.

**Table2.** Difference between experimental excitation energy of 102Sn low laying states and their corresponding calculated states using interaction fit1, fit2 and sn100pn.

|     | Exp-sn100pn | Exp-fit1 | Exp-fit2 |
| --- | --- | --- | --- |
| 0+  | 0 | 0 | 0 |
| 2+  | -0.211 | -0.044 | +0.020 |
| 4+  | -0.160 | +0.171 | +0.215 |
| 6+  | -0.011 | +0.170 | +0.017 |
| 8+  |  |  |  |
| 10+ |  |  |  |

Shell model calculations performed for the $^{104}$Sn isotope using the different interactions are represented in Fig. 4. As it is clearly seen, all the interactions reproduce correctly the spin and parity of all the levels in their correct order. For sn100pn interaction, except $6^+$ and $10^+$ states, the states are reproduced higher by about 200 keV compared to their corresponding experimental states. For fit1, $2^+$ is reproduced in its order and its position compared to the experiment. All other states are obtained lower to their experimental states by about 150 keV to 300 keV. On the other side for fit2 interaction, beside its reproduction of the correct order of spectrum levels, calculations give results in good agreement with the experimental spectrum for all states to an approximation of few tens of keV, except for $4^+$ which is obtained at level lower by about 180 keV.

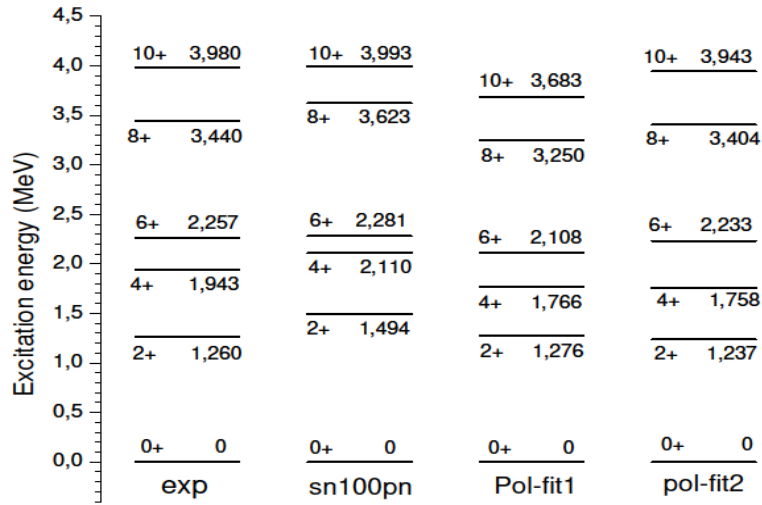

**Fig.4.** Experimental and calculated low laying states of 104Sn isotope.

**Table3.** Difference between experimental excitation energy of 104Sn low laying states and their corresponding calculated states using interaction fit1, fit2 and sn100pn.

|     | Exp-sn100pn | Exp-fit1 | Exp-fit2 |
| --- | --- | --- | --- |
| 0+  | 0 | 0 | 0 |
| 2+  | -0.234 | -0.016 | +0.023 |
| 4+  | -0.167 | +0.177 | +0.187 |
| 6+  | -0.024 | +0.149 | +0.024 |
| 8+  | -0.183 | +0.190 | +0.036 |
| 10+ | -0.013 | +0.310 | +0.037 |

The results of shell model calculations for $^{106}$Sn using the three interactions, fit 1, fit 2 and sn100pn, compared with its experimental spectrum are given in Fig. 5. All interactions reproduce in the correct order, spins and parities of all even states including the ground-state. For sn100pn interaction, all states are reproduced higher than their corresponding experimental levels. $2^+$ and $10^+$ states are higher by more than 200 keV, $4^+$ and $8^+$ are higher by about 170 keV and $6^+$ is reproduced higher by about 100 keV. For our two interactions, results are clearly better than those obtained by sn100pn interaction and are in good agreement with the experiment. Fit 1 reproduce three states among five to an approximation less than 100 keV. The two others are obtained less than their experimental corresponding's by about 150 keV. Likewise, fit 2 gives good results compared to the experiment. The three states, $2^+$, $6^+$ and $8^+$, are reproduced to an approximation less than 100 keV. Again, as it is for the two isotopes $^{102}$Sn and $^{104}$Sn, 4$^+$ is obtained less than the experiment by about 160 keV, while 10$^+$ is a bit higher by about 230 keV.

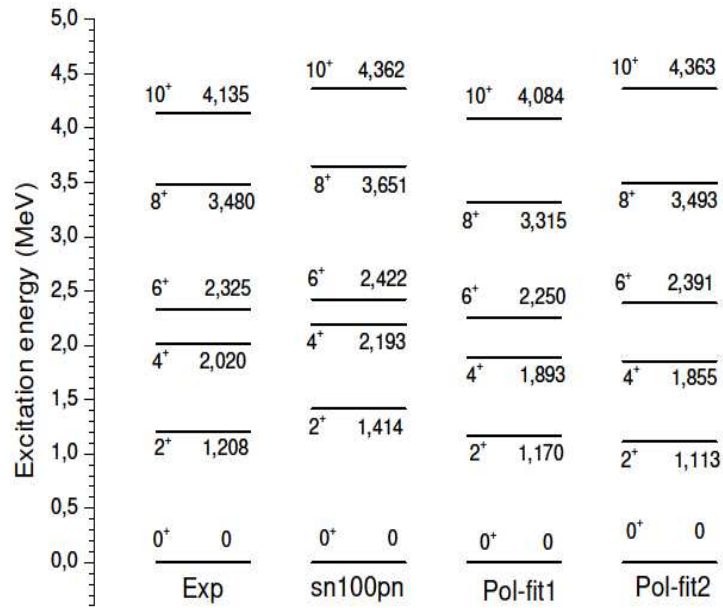

**Fig.5.** Experimental and calculated low laying states of 106Sn isotope.

**Table4.** Difference between experimental excitation energy of 106Sn low laying states and their corresponding calculated states using interaction fit1, fit2 and sn100pn.

|     | Exp-sn100pn | Exp-fit1 | Exp-fit2 |
| --- | --- | --- | --- |
| 0+  | 0 | 0 | 0 |
| 2+  | -0.206 | +0.038 | +0.095 |
| 4+  | -0.173 | +0.127 | +0.165 |
| 6+  | -0.097 | +0.075 | -0.066 |
| 8+  | -0.171 | +0.165 | -0.013 |
| 10+ | -0.227 | +0.051 | -0.228 |

**Odd tin isotopes: 103-107Sn**

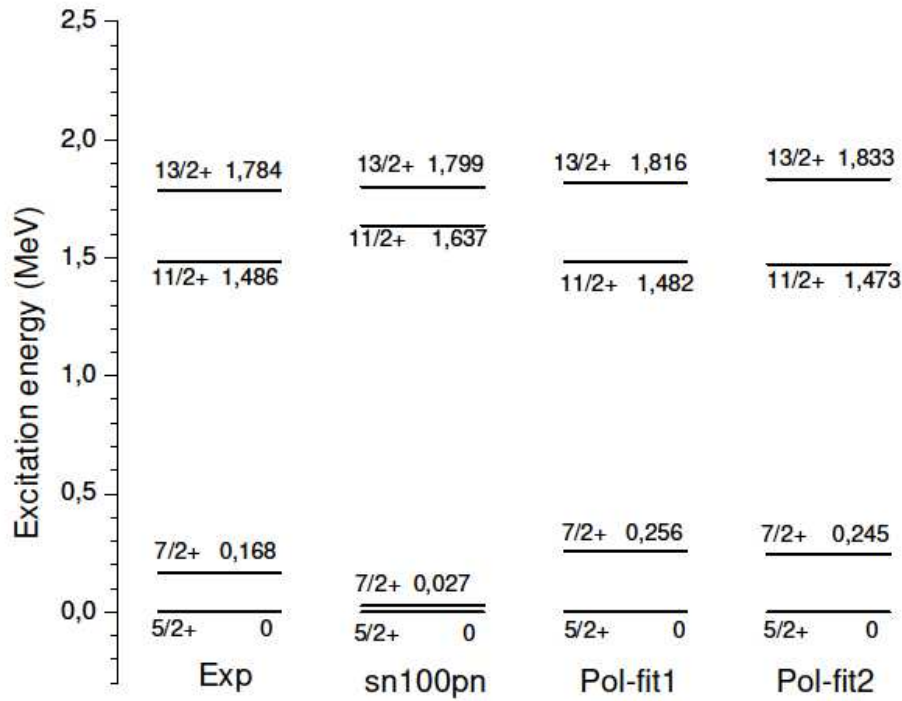

**Fig6:** Experimental and calculated low laying states of $^{103}$Sn isotope.

**Table5.** Difference between experimental excitation energy of 103Sn low laying states and their corresponding calculated states using interaction fit1, fit2 and sn100pn.

|       | Exp-sn100pn (MeV) | Exp-fit1 (MeV) | Exp-fit2 (MeV) |
|-------|-------------------|----------------|----------------|
| 5/2+  | 0                 | 0              | 0              |
| 7/2+  | +0.141            | -0.088         | -0.077         |
| 11/2+ | -0.151            | +0.004         | +0.013         |
| 13/2+ | -0.015            | -0.032         | -0.049         |

The results of shell model calculations for 103Sn using the three interactions, fit1 fit2 and sn100pn, compared with its experimental spectrum are given in fig6.

It is clear that there is a lack of experimental data for this nucleus, near the proton drip line, where only four levels figure in its experimental spectrum. Only two single particle states are in this spectrum where the ground state 5/2+ is separated to the first excited state 7/2+ by 0.168 MeV. The two other states, 11/2+ and 13/2+, are determined at a little bit higher levels at 1.486 MeV and 1.784 MeV respectively.

By using sn100pn interaction shell model calculations reproduce all states in their correct order. The ground state and 13/2+ states are in their positions by an approximation of few keV. The two other states 7/2+ and 11/2+ are reproduced less and more respectively by an approximation of 150 keV compared to their corresponding levels.

Results obtained by the two interactions fit1 and fit2 are much better than the corresponding shell model calculations using the well-known interaction sn100pn.
Beside the states are in their correct order, they are all reproduced in a very good agreement compared to the corresponding experimental levels by an approximation of few tens of keV. The two interactions are in the same order of approximation for this nucleus.

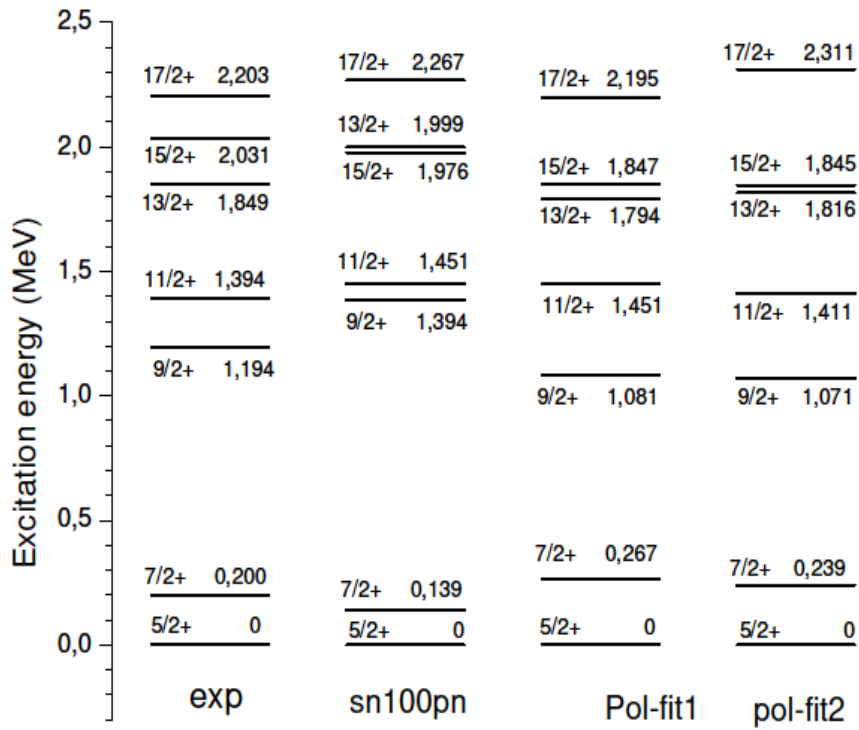

**Fig.7.** Experimental and calculated low laying states of 105Sn isotope.

**Table6.** Difference between experimental excitation energy of 105Sn low laying states and their corresponding calculated states using interaction fit1, fit2 and sn100pn.

|       | Exp-sn100pn (MeV) | Exp-fit1 (MeV) | Exp-fit2 (MeV) |
|-------|-------------------|----------------|----------------|
| 5/2+  | 0                 | 0              | 0              |
| 7/2+  | +0.061            | -0.067         | -0.039         |
| 9/2+  | -0.200            | +0.113         | +0.123         |
| 11/2+ | -0.057            | -0.057         | -0.017         |
| 13/2+ | -0.150            | +0.55          | +0.033         |
| 15/2+ | +0.055            | +0.184         | +0.186         |
| 17/2+ | -0.064            | +0.008         | -0.108         |

In Fig7 we represent results of shell model calculations using the sn100pn fit1 and fit2 interactions compared to the existing experimental spectrum for 105Sn nucleus. Results obtained by using the sn100pn interaction are in general in fair agreement with the experimental data. The majority of states are reproduced by an approximation of few tens of keV. The two states 9/2+ and 13/2+ are reproduced higher than their corresponding experimental levels by 200 keV and 150 keV respectively. The two levels 13/2+ and 15/2+ are reproduced in this Hamiltonian close to each other and in inverse order compared to the experimental spectrum for this nucleus.

For fit1 results, all states are reproduced in their correct order and in good agreement with their corresponding experimental levels by an approximation of few tens of keV, except 9/2+ and 15/2+ which are obtained at levels lower by about 110 keV and 180 keV respectively.

For fit2 results, beside the correct order of all states, three levels, 7/2+ 11/2+ and 13/2+, are reproduced in good agreement with the experimental spectrum by an approximation of few tens of keV. As in fit1 results, the states 9/2+ and 15/2+ are obtained at levels lower by about 120 keV and 180 keV respectively. 17/2+ is reproduced at level higher by more than 100 keV.

In general, even the obtained results for the three interactions are in fair agreement compared to the experiment, fit1 looks to be the best to reproduce the structure of 105Sn.

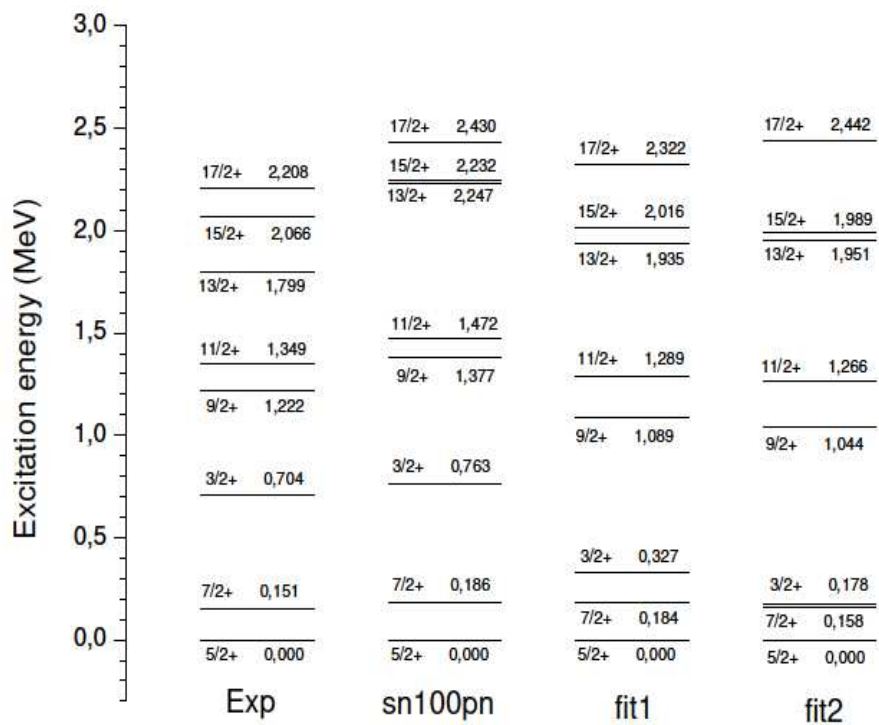

**Fig. 8.** Experimental and calculated low laying states of 107Sn isotope.

**Table7.** Difference between experimental excitation energy of 107Sn low laying states and their corresponding calculated states using interaction fit1, fit2 and sn100pn.

|          | Exp-sn100pn (MeV) | Exp-fit1 (MeV) | Exp-fit2 (MeV) |
|----------|-------------------|----------------|----------------|
| 5/2+     | 0                 | 0              | 0              |
| 7/2+     | -0.035            | -0.033         | -0.007         |
| 3/2+     | -0.059            | +0.377         | +0.528         |
| (3/2+)$_2$ |                 | +0.058         |                |
| (3/2+)$_3$ |                 | +0.021         |                |
| 9/2+     | -0.155            | +0.153         | +0.178         |
| 11/2+    | -0.123            | +0.060         | +0.083         |
| 13/2+    | -0.448            | -0.136         | -0.152         |
| 15/2+    | -0.166            | +0.050         | +0.077         |
| 17/2+    | -0.222            | -0.114         | -0.234         |
|          |                   |                |                |

The results of shell model calculations for 107Sn using the three interactions, fit1 fit2 and sn100pn, compared with its experimental spectrum are given in fig8.

Using the interaction sn100pn for this nucleus gives results with different accuracies. We can classify levels in three sets. The first set consists of the ground state 5/2+ and the first and the second excited states 7/2+ and 3/2+ respectively. These levels are reproduced in good agreement with the experimental data by an approximation of few tens of keV. The two states 9/2+ and 11/2+ consisting the second set are reproduced in a correct order but a bit higher by about 150 keV and 120 keV respectively compared to their corresponding experimental data. The third set of levels contains the three levels 13/2+ 15/2+ and 17/2+. All levels are reproduced higher with a less accuracy going from 160 keV to 450 keV.

In general for this interaction, only the first three levels including the ground state are reproduced in good agreement with the experiment.

For fit2 Hamiltonian results, beside the reproduction of the ground state, the three levels, 7/2+ 11/2+ and 15/2+, are reproduced in good agreement with the experimental spectrum by an approximation of few tens of kev. In contrast, shell model calculations using this interaction reproduce three other levels, 9/2+ 13/2+ and 17/2+, with a less accuracy by approximation from 150 KeV to 230 keV. It is clear that this interaction failed to reproduce the single particle state 3/2+ which is reproduced in a level very low than its corresponding experimental position by more than 520 keV.

For fit1 results, all states are reproduced in good agreement with their corresponding experimental levels. Four states, are reproduced by an approximation of few tens of keV. The other three states are reproduced in fair agreement with approximations between 100 KeV and 150 KeV.

For 3/2+, shell model calculations using fit1 interaction predict three levels for this state. The first level is predicted much lower by an approximation of 380 KeV than the corresponding experimental one. In addition, these calculations predict this state in two other levels close to each other, and in very good agreement with the experiment by an approximation of few tens of keV. Results of fit2 interaction don't predict a second level for 3/2+ state and the predicted one is very far to the experiment.

For this nucleus, fit1 is the best to reproduce the experimental spectrum compared to fit1 and sn100pn interactions.

**Conclusion**

In this paper, we have used a phenomenological approach based on the experimental data to derive the single particle states in 101Sn. For this purpose, and based on the trends evolution of the experimental data, we used two second order polynomial fits fit1 and fit2 for each set of experimental data belonging to single particle states d5/2 g7/2 d3/2 s1/2 and h11/2 in light odd Sn isotopes. By extrapolating curves down forward to proton limit of existence, we obtained the excitation energies of these states in 101Sn isotope, a nucleus with one neutron outside the doubly magic core 100Sn. Subsequently, we defined single particle energies needed in shell model calculation.

Shell model calculations of nuclear spectra of even and odd tin 102-107Sn isotopes have been carried out using the obtained interactions and the well-known interaction sn100pn. All results are compared with the experimental spectrum for each nucleus.

For even tin 102-106Sn isotopes, fit2 interaction is the best to give results in good agreement with the experimental data compared to results obtained with fit1 and sn100pn interactions.

In contrast, the results of the shell model calculations performed with fit1 interaction for odd tin 103-107Sn isotopes are in in full agreement with the experimental data and better than results obtained using fit2 and sn100pn interactions.

**References**

[1] A. Holt et al., The structure of neutron deficient Sn isotopes, Nucl. Phys. A 570, 137 (1994).
[2] A. Insolia et al., Microscopic structure of Sn isotopes, Nucl. Phys. A 550, 34 (1992).
[3] Z. H. Sun et al., Effective shell-model interaction for nuclei southeast of 100Sn, Phys. Rev. C 104, 064310 (2021).
[4] A. Yakhelef and A. Bouldjedri, Shell model calculation for Te and Sn isotopes in the vicinity of 100Sn, in The 8th International Conference on Progress in Theoretical Physics(ICPTP 2011), 23–25 October 2011, Constantine, Algeria,edited by N. Mebarki, J. Mimouni, N. Belaloui, and K. Ait Moussa, AIP Conf. Proc. No. 1444 (AIP, New York, 2012), p. 199.
[5] B. A. Brown, N. J. Stone, J. R. Stone, I. S. Towner, and M. Hjorth-Jensen, Magnetic moments of the 2+1 states around 132Sn, Phys. Rev. C 71, 044317 (2005).
[6] A. Hosaka et al., G-matrix effective interaction with the paris potential, Nucl. Phys. A 444, 76 (1985).
[7] T. Trivedi et al., Shell model description of 102−108Sn isotopes, Int. J. Mod. Phys. E 21, 1250049 (2012).
[8] G. A. Leander et al., Single-particle levels in the doubly magic 132Sn and 100Sn nuclei, Phys. Rev. C 30, 416 (1984).
[9] T. Engeland et al., Large shell model calculations with realistic effective interaction, Phys. Scr. T56, 58 (1995).
[10] F. Andreozzi, L. Coraggio, A. Covello, A. Gargano, T. T. S. Kuo, Z. B. Li, and A. Porrino, Realistic shell-model calculations
for neutron deficient Sn isotopes, Phys. Rev. C 54, 1636 (1996).
[11] N. Sandulescu et al., Microscopic description of light Sn isotopes, Nucl. Phys. A 582, 257 (1995).
[12] H. Grawe et al., In-beam spectroscopy of exotic nuclei with OSIRIS and beyond, Prog. Part. Nucl. Phys. 28, 281 (1992).
[13] R. Schubart *et al.*, Shell model structure at $^{100}$Sn—The nuclides 98Ag, 103In, and 104-105Sn, Z. Phys. A 352, 373 (1995).
[14] M. G. Mayer, On closed shells in nuclei, Phys. Rev. **74**, 235 (1948).
[15] I. Talmi, Fifty years of the shell model – The quest for the effective interaction, Adv. Nucl. Phys. **27**, 1 (2003).

[16] E. Caurier, G. Martínez-Pinedo, F. Nowack, A. Poves, and A. P. Zuker, The shell model as a unified view of nuclear structure, Rev. Mod. Phys. **77**, 427 (2005).

[17] B. A. BROWN, The nuclear shell model towards the drip lines, Prog. Part. Nucl. Phys 47, 517 (2001).

[18] B. A. Brown, W. D. M. Rae, The shell-model code NuShellX@MSU, Nucl. Data Sheets **120**, 115 (2014).

[19] Nushell@MSU, B. A. Brown and W. D. M. Rae, MSU-NSCL report (2007).

[20] C. Lanczos: An Iteration Method for the Solution of the Eigenvalue Problem of Linear Differential and Integral Operators, J. Res. Nat. Bur. Stand. 49, 255 (1950)

[21] B. A. Brown, N. J. Stone, J. R. Stone, I. S. Towner, and M. Hjorth-Jensen, Magnetic moments of the $2^+_1$ states around $^{132}$Sn, Phys. Rev. C **71**, 044317 (2005).

[22] BNL Evaluated Nuclear Structure Data File (ENSDF),http://www.nndc.bnl.gov/ensdf/.

[23] D. Seweryniak *et al.*, Phys. Rev. Lett. **99**, 022504 (2007).

[24] I. G. Darby *et al.*, Phys. Rev. Lett. **105**, 162502 (2010).